\documentclass{article}
\usepackage{spconf,amsmath,graphicx}

\usepackage{url}
\usepackage{graphicx}
\usepackage{mathtools}
\usepackage{makecell}
\usepackage{color, soul}
\usepackage{multirow}

\usepackage{xcolor,colortbl}
\definecolor{Gray}{gray}{0.90}
\definecolor{Gray1}{gray}{0.95}

\usepackage{acro}

\DeclareAcronym{asic}{short = ASIC , long  = Application-Specific Integrated Circuit}
\DeclareAcronym{amt}{short = AMT , long  = Adaptive Multiple Transform}

\DeclareAcronym{brams}{short = BRAMS , long  = Block RAMS}

\DeclareAcronym{dct}{short = DCT , long  = Discrete Cosine Transform}
\DeclareAcronym{dst}{short = DST , long  = Discrete Sine Transform }
\DeclareAcronym{dc}{short = DC , long  = Design Compiler}

\DeclareAcronym{fpga}{short = FPGA , long  = Field-Programmable Gate Array}
\DeclareAcronym{fpgas}{short = FPGAs , long  = Field-Programmable Gate Array}
\DeclareAcronym{hevc}{short = HEVC , long  = High Efficient Video Coding}

\DeclareAcronym{ietr}{short = IETR , long  = Institut d'\'Electronique et de T\'el\'ecommunications de Rennes}
\DeclareAcronym{idst}{short = IDST , long  = Inverse DST }
\DeclareAcronym{idct-ii}{short = IDCT-II , long  = Inverse DCT-II}

\DeclareAcronym{jvet}{short = JVET , long  = Joint Video Experts Team}

\DeclareAcronym{mts}{short = MTS , long  = Multiple Transform Selection }
\DeclareAcronym{mpeg}{short = MPEG , long  = Motion Picture Experts Group}

\DeclareAcronym{ram}{short = RAM , long  = Random-Access Memory}
\DeclareAcronym{rom}{short = ROM , long  = Read-Only Memory}
\DeclareAcronym{rm}{short = RM , long  = Regular Multiplier}
\DeclareAcronym{mcm}{short = MCM , long  = Multiple Constant Multiplier}

\DeclareAcronym{tsmc}{short = TSMC , long  = Taiwan Semiconductor Manufacturing Company}

\DeclareAcronym{vvc}{short = VVC , long  = Versatile Video Coding}
\DeclareAcronym{vceg}{short = VCEG , long  = Video Coding Experts Group}

\title{Lightweight Hardware Implementation of VVC Transform Block for ASIC Decoder}
\name{I. Farhat$^{\dagger \star}$, W. Hamidouche$^{\dagger}$, A. Grill$^{\star}$, D. Menard$^{\dagger}$ and O. D\'eforges$^\dagger$}
\address{$^\dagger$ Univ Rennes, INSA Rennes, CNRS, IETR - UMR 6164, Rennes, France. \\ 
	$^\star$ VITEC, 99 rue Pierre Semard, 92320 Chatillon, France. \\E-mail: ibrahim.farhat@vitec.com, adrien.grill@vitec.com,  whamidou@insa-rennes.fr
}

\begin{document}
\ninept
\maketitle
\begin{abstract}
\ac{vvc} is the next generation video coding standard expected by the end of 2020. Compared to its predecessor, \ac{vvc} introduces new coding tools to make compression more efficient at the expense of higher computational complexity. This rises a need to design an efficient and optimised implementation especially for embedded platforms with limited memory and logic resources. One of the newly introduced tools in \ac{vvc} is the \ac{mts}. This latter involves three \ac{dct}/\ac{dst} types with larger and rectangular transform blocks. In this paper, an efficient hardware implementation of all DCT/DST transform types and sizes is proposed. The proposed design uses 32 multipliers in a pipelined architecture which targets an ASIC platform. It consists in a multi-standard architecture that supports the transform block of recent MPEG standards including AVC,  HEVC and VVC. The architecture is optimized and removes unnecessary complexities found in other proposed architectures by using regular multipliers instead of multiple constant multipliers. The synthesized results show that the proposed method which sustain a constant throughput of two pixels/cycle and constant latency for all block sizes
can reach an operational frequency of 600 Mhz enabling to decode in real-time 4K videos at 48 fps. 
\end{abstract}

\acresetall

\section{Introduction}
The next generation video coding standard named \ac{vvc} is under development by the \ac{jvet}, established by \ac{mpeg} and \ac{vceg}. The \ac{vvc} standard, expected by the end of 2020, introduces several new coding tools enabling up to 40\%  of coding gain beyond \ac{hevc} standard~\cite{JVET-O0003, PCS-2019-VVC-QE}. One of the newly introduced tools is the \ac{mts} which involves three transform types including \ac{dct} type II (DCT-II), \ac{dct} type VIII (DCT-VIII)  and \ac{dst} type VII (DST-VII) with block sizes that can reach 64 $\times$ 64 for DCT-II and 32 $\times$ 32 for DCT-VIII/DST-VII. The use of \ac{dct}/\ac{dst} families gives the ability to apply separable transforms, the transformation of a block can be applied separately in horizontal and vertical directions. If the DST-VII horizontal direction is selected, VVC enables the use of DCT-VIII or DST-VII for the vertical direction, and vice versa, unlike DCT-II which is applied for both directions when it is selected by the encoder. 


In this paper, two hardware implementations with two different architectures of the MTS transforms are proposed. The first uses Hcub Multiplierless \ac{mcm} algorithm \cite{mcmarch} and the second relies on \ac{rm} to compute the transform. These implementations support 1-D \ac{idct-ii} of orders from 4 to 64 and IDST-VII/IDCT-VIII of orders from 4 to 32. The 2-D transform can be performed using two 1-D transforms by adding an intermediate transpose memory in a folded architecture. Both modules can support the transform of the recent \ac{mpeg} video coding standards including AVC, HEVC and VVC. The main consideration of these implementations is to conserve a fixed throughput of 2 pixels/cycle and a fixed system latency for all transform sizes and types. This enables accurate prediction of the process performance while facilitating chaining between transform blocks. To the best of our knowledge, this is the first implementation that includes all VVC-MTS transforms with size up to 64 for the DCT-II. 
Synthesis results using Synopsys \ac{dc} tool, show that the \ac{rm} architecture consumes 63\% less gates than the \ac{mcm} one. For our multi-standard decoder, we adopt the \ac{rm} architecture due to the significant surface gain.

The rest of this paper is organized as follows. Section~\ref{sec:related-works} presents the background of the MTS kernels and the existing works related its hardware implementations. In Section~\ref{sec:proposed-hard-impl}, the detailed hardware implementations of the \ac{mcm} and \ac{rm} architectures are presented. Experimental setup, test conditions and results are described in Section~\ref{sec:expremental-set}, along with a comparison with the state of art works. 
Finally, Section~\ref{sec:conclusion} concludes this paper.

\section{Related Works}
\label{sec:related-works}
\subsection{Background of the MTS}
The HEVC standard is based on the DCT type II as the main transform function and the DST of type VII applied only for Intra blocks of size 4 $\times$ 4. In VVC, the MTS  scheme can be used for coding both Intra and Inter blocks.  

The basis functions of the three transforms considered in the MTS module are expressed by Equations (\ref{Equ:dct-2}), (\ref{Equ:dst-7}) and (\ref{Equ:dct-8}) for DCT-II $C_2$, DST-VII $S_7$ and DCT-VIII $C_8$, respectively. MTS extends the use of the DST-VII/DCT-VIII for blocks of sizes 8$\times$8, 16$\times$16 and 32$\times$32 including all possible asymmetric blocks, and also considers the 64$\times$64 block size for the DCT-II.
\begin{equation}
C_{2\, i, j} = w_i \sqrt{\frac{2}{N}} \cos \left ( \pi i \frac{(2j+1)}{2N}  \right ) ,
\label{Equ:dct-2}
\end{equation}
with $w_i=\left\{ \begin{array}{cc}
\sqrt{\frac{1}{2}} & i=0 \\ 
1 & i\in \{1, \dots ,N-1 \}\end{array}
\right.$.

\begin{equation}
S_{7\, i,j}  = \sqrt{\frac{4}{2N+1}} \sin \left (\pi \frac{(2i+1) (j+1)}{2N+1} \right).
\label{Equ:dst-7}
\end{equation}

\begin{equation}
C_{8\, i, j} = \sqrt{\frac{4}{2N+1}} \cos \left (\pi  \frac{(2i+1) (2j+1)}{4N+2} \right ). 
\label{Equ:dct-8}
\end{equation} 

The 2D forward transform of an input block $X$ of size $M \times N$ is computed by 
\begin{equation} 
Y =  B_V \cdot X \cdot B_H^T 
\end{equation}
where $B_H^T$ is a matrix of size $N \times N$ of horizontal transform coefficients, $B_V$  is a matrix of size $M \times M$ of vertical transform coefficients.  
The inverse 2D transform is computed by Equation (\ref{equ:inv2D})
\begin{equation} 
\hat{X} =  B_V^T \cdot \tilde{Y} \cdot B_H 
\label{equ:inv2D}
\end{equation}



Finally, to further decrease the computational complexity of the transform module, high frequency transform coefficients are zeroed out for the transform blocks of sizes equal to 64 for DCT-II and 32 for transform types DST-VII/DCT-VIII. Therefore, only lower-frequency coefficients are retained. 

\subsection{Existing Hardware Implementations}

Several DCT-II hardware implementations have been proposed in the literature, as it is the main transform used in the previous video coding standards. Shen {\it et al.} \cite{6298499} have presented a unified 4/8/16 and 32-point DCT-II targeting ASIC platform. They used \ac{mcm} for sizes 4 $\times$ 4 and 8 $\times$ 8 and \ac{rm} for 16 $\times$ 16 and 32 $\times$ 32 transform blocks. Moreover, the proposed architecture is a multi-standard supporting five video standards including AVC, HEVC, AVS, VC1 and MPEG-2/4 with a fixed throughput of 4 pixels/cycle. 

Recently, several works on hardware implementation of the VVC transform have been published. Kammoun {\it et al.}~\cite{Kammoun4x4} proposed a 1-D hardware implementation of the \ac{amt} including five transform types DCT-II, DST-I, DST-V, DST-VII and DCT-VIII for only size 4$\times$4 using adders and shifts instead of regular multipliers. Although this work presents a hardware implementation for all transform types (including the MTS types), it only supports 4 $\times$ 4 block sizes. Mert~{ \it et al.}~\cite{Can-mert} proposed another hardware implementation supporting also all transform types for sizes 4 $\times$ 4 and 8 $\times$ 8. This solution investigated two hardware methods with a fixed 8 pixels/cycle throughput. The first one uses separate data paths and the second one considers re-configurable data paths for all 1-D transforms. This solution is limited to block size of 8 $\times$ 8, knowing that the transform of large block size (16$\times$16, 32$\times$32 and 64$\times$64) is much more complex and requires more resources. 

Garrido {\it et al.}~\cite{Garrido} have proposed a pipelined 1-D hardware implementation of the \ac{amt} of size 4 $\times$ 4, 8 $\times$ 8 ,16 $\times$ 16 and 32 $\times$ 32. It includes the five transform types within the JEM software \cite{JVET-B0022}. The proposed design has a fixed throughput of 4 pixels/cycle on a fully pipelined architecture. The design  relies on two 1D implementations connected with a transpose memory. Garrido { \it et al.} in \cite{8698857} extended their work to support a 2D transform. The synthesis results targeting multiple state of the art \ac{fpgas} platforms show that the design can support in the best scenario 4K video resolution at 23 frames per second. Although the work in \cite{Garrido} supports all VVC transform types, it does not support blocks of size 64 $\times$ 64 which significantly increases the complexity. Kammoun { \it et al.} \cite{Kammoun} proposed another hardware implementation for the same transform types and sizes. It implements the transformation using IP cores multiplier taking advantage of DSPs of the FPGA device. The work in \cite{Kammoun} proposes a 2-D hardware implementation. Although it includes 2-D implementation, blocks of sizes 64 $\times$ 64 are also not supported. Authors in \cite{ledernier} proposed a 2D hardware architecture for DCT-VIII and DST-VII of VVC. They implement all DST-VII DCT-VIII transform sizes using adders and shifts. Although it is the first architecture that includes asymmetric blocks, it does not include the DCT-II.

This paper proposes a comparison of two 1-D hardware architectures for 4/8/16/32/64 1-D MTS transform cores. The first architecture uses adders and shifts for all types and sizes to perform a 1-D transform operation, while the second one relies on 32 regular multipliers to perform the required multiplications. \ac{rom} is used to store coefficients for the \ac{rm} architecture. Up the best of our knowledge, this is the first implementation of a \ac{vvc} \ac{mts} architecture supporting all transform types and sizes.

\section{Proposed hardware implementations}
\label{sec:proposed-hard-impl}

\begin{table}[!b]
\caption{MTS design interface}
\centering
\tabcolsep=0.10cm
\begin{tabular}{|l|c|c|c|}
	\hline
	Signal & I/O & \#BITS & Description \\
	\hline
	\hline
	$clk$ & I& 1& System Clock\\	
	\hline
	$rst\_n$ & I & 1 & System reset, active low\\
	\hline
	$input\_enable$ & I& 1& \makecell{ Activation pulse to start}\\
	\hline
	$avc\_vvc$ & I& 1& \makecell{ Video standard: 0, avc;\\ 1, hevc or vvc}\\
	\hline
	$tr\_type$ & I& 2& \makecell{Transform type: 0, DCT-II;\\ 1, DCT-VIII; 2,DST-VII}\\
	\hline
	$tr\_size$ & I & 3 & \makecell{Transform size: 0:4-pnt; 1:8-pnt;\\ 2:16-pnt; 3:32-pnt;4:64-pnt}\\
	\hline
	$tr\_dir$	& I & 1 & \makecell{Transform direction : \\ 0:Horizontal; 1: Vertical}\\
	\hline
	$data\_in$& I & 2$\times$ $N_{bi}$ &Input data \\
	\hline
	$data\_enable$ & O & 1 & \makecell{Activation pulse to indicate \\ end of N-point}\\
	\hline
	$data\_out\_inter$ & O & 2$\times$$N_{bi}$ & \makecell{Intermediate output data }\\
	\hline
	$data\_out\_fin$ & O & 2$\times$$N_{bo}$ & \makecell{Final result }\\
	\hline
\end{tabular}

\label{tab:MTS-interface}
\end{table}

In this section, we present a brief description of the two proposed hardware designs for the 1-D MTS transforms. For both architectures, the constraint is to sustain a fixed throughput of 2 pixels/cycle and a fixed latency regardless of the block size. This choice enables the transformation to process a 1-D horizontal or vertical 4, 8, 16, 32 and 64-point sizes in 8, 32, 128, 512 and 2048 clock cycles, respectively. Moreover, we introduced a delay line at the output to provide a fully pipelined structure. The size of the delay line has been determined according to the throughput and the largest transform block latency. This static latency enables a smooth chaining between processing engines of the decoder.

The MTS processor interfaces for both designs are summarized in table \ref{tab:MTS-interface}. A positive pulse in $input\_enable$ launches the transform process, with the size, type and transform direction are defined in the $tr\_size$, $tr\_type$ and $tr\_dir$ input signals, respectively. Following the $input\_enable$ the $data\_in$ signal will carry two $N_{bi}$ coefficients at the next clock cycle. After the computation, outputs can be read from the $data\_out\_inter$ or $data\_out\_final$ signals, depending on the transform direction, at a speed of two $N_{bi}$ pixels each clock cycle. For both implementations that integrate data loading and pipeline stages, the design starts generating the result of 1-D row/columns DCT/DST after a fixed system latency. It then continues generating the outputs every clock cycle without any stalls.

\subsection{Multiplierless Architecture}

The \ac{mcm} architecture is a design in which multipliers are replaced by adders and shifts. It is widely used and for small block sizes, it has been proven to be more efficient than a regular multiplier architecture. However, VVC introduces new transform types with larger 64 $\times$ 64 block size. These new transforms increase the number of coefficients and the number of possible multiplications. For that \ac{rm} architecture is also reconsidered in this paper. 

\begin{figure}[hbt!] 
\centering
\includegraphics[width=1\linewidth]{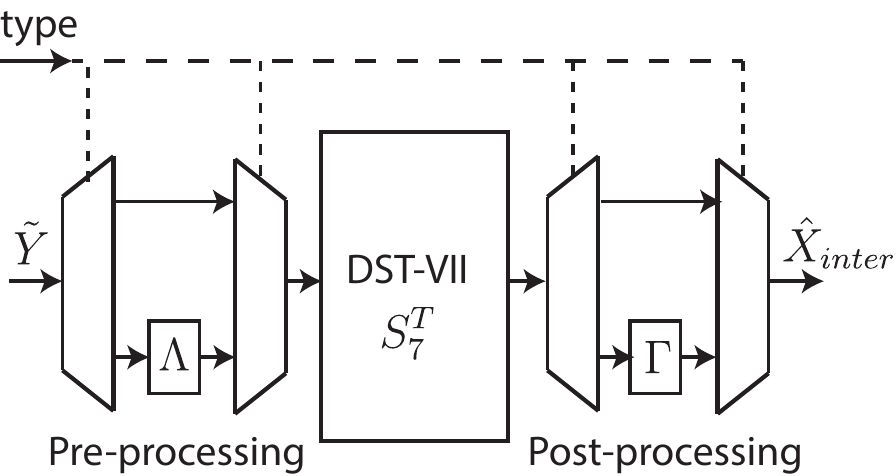}
\centering
\caption{N-point IDCT-VIII/IDST-VII hardware design using IDST-VII as kernel transformation, pre-processing and post-processing blocks are used when IDCT-VIII type is active.}
\label{fig:dst7-dct8}
\end{figure}

The \ac{mcm} architecture contains five separable modules, one module for the N-point IDCT-II in a unified architecture and four independent modules for 4-point, 8-point, 16-point and 32-point IDCT-VIII/IDST-VII. We take advantage of the butterfly decomposition to regroup all IDCT-II sizes (4-point, 8-point, 16-point, 32-point, 64-point) into one unified architecture. For the IDCT-VIII and IDST-VII which do not have any specific decomposition, we created one module for each block size including 4-point, 8-point, 16-point and 32-point. The relationship between IDCT-VIII and IDST-VII enables to use only the IDST-VII kernel to compute both transformations by adding two stages for preprocessing and postprocessing as shown in Figure~\ref{fig:dst7-dct8}. Equation (\ref{Eq9}) computes the IDCT-VIII $C_8^T$ from the IDST-VII $S_7^T$ using pre-processing $\Lambda$ and post-processing $\Gamma$ matrices.  
\begin{equation}
{C}^T_{8}  = \Lambda \cdot S^T_{7} \cdot \Gamma,   
\label{Eq9}
\end{equation}
where $\Lambda$ and $\Gamma$ are permutation and sign changes matrices computed by Equations~(\ref{Eq5}) and ~(\ref{Eq6}), respectively.
\begin{equation}
\Lambda_{i,j}= \left\{ \begin{array}{cc}1, & \textit{ if } j=N-1-i, \\ 0, & \text{ otherwise }, \end{array}\right. \\
\label{Eq5}
\end{equation}
\begin{equation}
\Gamma_{i,j}=\left\{ \begin{array}{cc} (-1)^{i}, & \textit{ if } j=i,  \\ 0, & \text{ otherwise }, \end{array}\right. \\ 
\label{Eq6}
\end{equation}
with $i, j \in \{0, 1, \dots,  N-1\}$ and $N \in \{4, 8, 16, 32 \}$.

Therefore, the IDCT-VIII is computed only by inverting the input order using a pre-processing stage and assigning the appropriate outputs signs using a post-processing stage.

\begin{figure}[t] 
\centering
\includegraphics[width=0.9\linewidth]{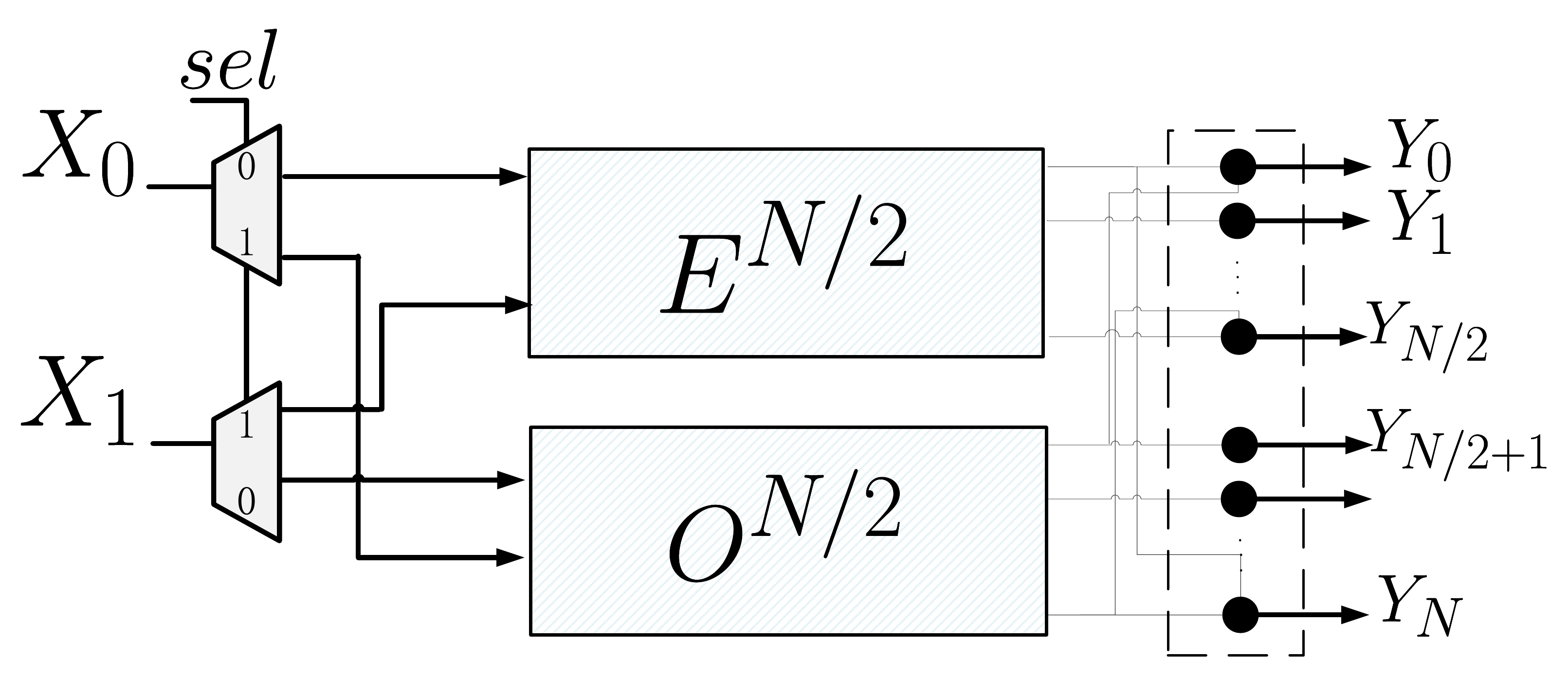}
\caption{General structure of the order-64 inverse DCT-II. Blocks $O^{N/2}$ corresponds to odd decomposition and $E^{N/2}$ to the even decomposition of $[C_2^N]^T$. $X_0$ and $X_1$ denote the input samples, while $Y_i$ denote the outputs.}
\label{dct-2-butterfly}
\end{figure}

Several IDCT-II hardware implementations have been proposed in the literature as it is the classic transform used in the previous video coding standards. However, none of the existing implementations supports the size of 64-point IDCT-II, until now this is the first proposed solution based on a constant multiplier architecture that reaches the size 64. Figure~\ref{dct-2-butterfly} shows the unified N-point hardware design for the IDCT-II. The proposed implementation relies on the state-of-the-art butterfly architectures and includes the IDCT-II of order 64. 

\begin{figure}[hbt!] 
\centering
\includegraphics[width=0.9\linewidth]{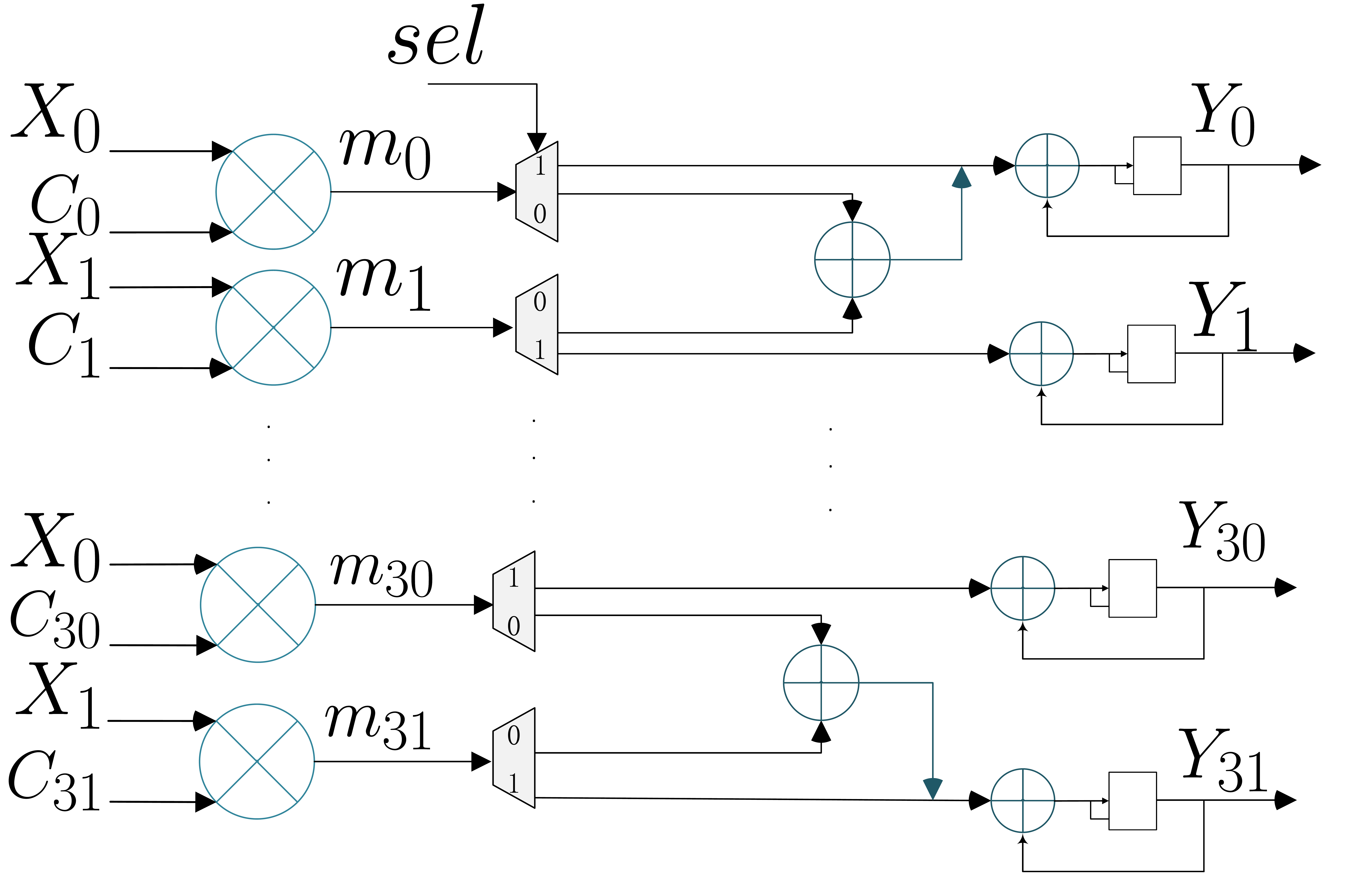}
\centering
\caption{\ac{rm} architecture, $X_0$ and $X_1$ are the input samples, $C_i$ is the transform coefficient, $m_i$ is a multiplier and $Y_i$ represents the output}
\label{fig:shared_mult}
\end{figure}

\begin{table}[!h]
\renewcommand{\arraystretch}{1.1}
\caption{Synthesis results for both ASIC and FPGA platforms}
\centering
\begin{tabular}{|l|l|c|l|c}
	\hline
 	& & MCM arch. & RM arch.\\
	\hline
	\hline
\multirow{5}{*}{ASIC}	& Frequency (Mhz) & 600 & 600 \\
	\cline{2-4}
	& Combinational area & 221659 & 59588 \\
	\cline{2-4}
	& Noncombinational area & 27860 & 32413 \\
	\cline{2-4}
	& Buf/Inv area & 16796 & 4848 \\
	\cline{2-4}
	& Total area & 266315 & 96849 \\
	\hline
	\hline
\multirow{4}{*}{FPGA}	& Frequency (Mhz) & 139 & 165 \\
\cline{2-4} 
	& ALMs (/427200) & 51182 (12\%) & 9723 (2\%) \\
	\cline{2-4}
	& Registers & 16924 & 14368 \\
	\cline{2-4}
	& DSP blocks(/1518) & 0 (0\%) & 32 (2\%) \\
	\hline 
\end{tabular}
\label{tab:synASIC}
\end{table}

\begin{table*}[!ht]
\centering
\renewcommand{\arraystretch}{1.1}
\caption{Comparison of different hardware transform designs}
\vspace{1mm}
\begin{tabular}{|p{0.86in}|p{0.7in}|p{0.85in}|p{0.9in}|p{0.9in}|p{1.7in}|} \hline 
Solutions&  \centering{Fan {\it et al.}~\cite{ledernier}} & \centering{Garrido {\it et al.}~\cite{8698857}} & Mert {\it et al.} \centering{~\cite{Can-mert}} & Kammoun {\it et al.} \centering{~\cite{Kammoun}}& \centering{Proposed RM architecture}\tabularnewline \hline 
\rowcolor{Gray1} Technology  &  \centering{ASIC 65 nm} & \centering{20 nm ME}&  \centering{ASIC 90 nm} & \centering{ME 20 nm FPGA}& \centering{\begin{tabular}{p{0.75in} | p{0.7in}}ASIC 28 nm  &  ME FPGA \end{tabular}}  \tabularnewline  
 Gates/ALMs   &  \centering{496400} & \centering{1312}&  \centering{ 417000} & \centering{$133017$}& \centering{\begin{tabular}{p{0.75in} | p{0.7in}} \centering{96849}   & \centering{9723}   \end{tabular}} \tabularnewline 
 \rowcolor{Gray1} Registers   &  \centering{--} & \centering{3624}&  \centering{--} & \centering{--}& \centering{\begin{tabular}{p{0.75in} | p{0.7in}} \centering{--}  &  \centering{14368}  \end{tabular}}   \tabularnewline
 DSPs   &  \centering{--} & \centering{32} &  \centering{--} & \centering{1561}& \centering{\begin{tabular}{p{0.75in} | p{0.7in}}\centering{--}  &  \centering{32} \end{tabular}} \tabularnewline 
\rowcolor{Gray1} Frequency (Mhz)   &  \centering{250} & \centering{{458.72}}&  \centering{160} & \centering{{$147$}}& \centering{\begin{tabular}{p{0.75in} | p{0.7in}}\centering{600}  &  \centering{165} \end{tabular}} \tabularnewline 
Throughput (fps) &  \centering{--} & \centering{3840 $\times$ 2160p23}&  \centering{7680 $\times$ 4320p39} & \centering{1920 $\times$ 1080p50}& \centering{\begin{tabular}{p{0.75in} | p{0.7in}}3840$\times$2160p30  &  1920$\times$1080p50 \end{tabular}} \tabularnewline 
\rowcolor{Gray1}Memory  &  \centering{$-$} & \centering{41 $\times$ 21 Kbits}&  \centering{--} & \centering{--}& \centering{--} \tabularnewline 
Transform size  &  \centering{ 4, 8, 16, 32}  &   \centering{4, 8, 16, 32}   &  \centering{ 4, 8} & \centering{ 4, 8, 16, 32} &  \centering{4, 8, 16, 32, 64}  \tabularnewline  
\rowcolor{Gray1}\multirow{2}{*}{Transform type} & \centering{ {\ac{dct}-II/VIII, \ac{dst}-VII} } & \centering{ \ac{dct}-II/VIII, \ac{dst}-VII}  &  \centering{ \ac{dct}-II/VIII, \ac{dst}-VII} & \centering{  \ac{dct}-II/V/VIII, \ac{dst}-I/VII } & \centering{\multirow{2}{*}{\ac{dct}-II/VIII, \ac{dst}-VII}}\tabularnewline 
\hline 
\end{tabular}
\label{comparison2D}
\end{table*}

\subsection{Regular Multipliers Architecture}

In this section, the \ac{rm} architecture is investigated. The architecture of 4/8/16/32-point 1D IDCT-II/VIII, IDST-VII and 64-point 1D IDCT-II uses 32 shared multipliers. Thirty-two is the maximum number of multiplications needed to get an output rate of 2 pixels/cycle. This number is bounded by the odd decomposition of the 64$\times$64 IDCT-II transform matrix and the 32$\times$32 IDST-VII/IDCT-VIII transform matrices. For large block sizes and by taking advantage of the zeroing, we can process one pixel/cycle at the input to get a 2 pixel/cycle at the output. In the case of 64-point IDCT-II, the size of the input vector is 32 and the output vector size is 64. For the 32-point IDCT-VIII/IDST-VII, the size of the input and output vectors are 16 and 32, respectively. In both cases, the output vector is twice the size of the input vector, and thanks to this, we can lower the input rate to one pixel per cycle. 

Figure~\ref{fig:shared_mult} shows the architecture of the hardware module using 32 \ac{rm}s referred to $m_i$. $X_0$ and $X_1$ are the input samples. Using the zeroing for large block sizes, $X_0$ and $X_1$ interfaces will carry the same input sample and $sel$ signal is disabled, otherwise they carry two different samples and $sel$ is enabled. The input samples are then multiplied by the corresponding transform coefficients $C_i$ $(i \in {1 .. N})$. $C$ represents a line from the transform matrix. Each $X_i$ is multiplied by its corresponding $C_i$ coefficient. The result is then accumulated at the output vector $Y_i$ using the adders and the feedback lines as shown in Figure~\ref{fig:shared_mult}.

The transform coefficients are stored in a \ac{rom}. The total memory size is 17408-bits which corresponds to 68 columns of 256 bit-depth (68$\times$256). The \ac{rom} stores the coefficients of the 64-point IDCT-II, 32-point, 16-point, 8-point and 4-point IDST-VII matrix coefficients. The 64-point IDCT-II is decomposed using its butterfly structure, and the resultant sub matrices are stored. In fact, one sub matrix is replicated to respect the output rate. The relationship between IDCT-VIII and IDST-VII enables us to compute both transforms using one kernel. Thus, we store only the IDST-VII transform matrices.

\section{Experimental Results}
\label{sec:expremental-set}

VHDL hardware description language is used to implement both proposed designs. A state-of-the-art logic simulator \cite{RiviraPro} is used to test the functionality of the 1D transform module. The test strategy is as follows.  First a set of $10^5$ pseudo-random input vectors have been generated and used as test patterns. Second, a software implementation of the inverse transforms has been developed, based on the transform procedures used in VTM 6.0 \cite{JVET-O0003}. Using self-check techniques, the bit accurate test-bench compares the simulation results with those obtained using the reference software implementation. 

The proposed design supports three different video standards including AVC/H.264, HEVC/H.265 and the emerging VVC/H.266 standard. The \ac{mcm} 1-D architecture MTS core works at 600Mhz with 249K cell area, while the \ac{rm} architecture operates at 600Mhz with 93K cell area and 17408 bits of ROM used to store transforms coefficients. 


The 1D-MTS module has been Synthesised by \ac{dc} with \ac{tsmc} 28nm standard cell library targeting ASIC platform.  Table~\ref{tab:synASIC} gives the synthesis results for both architectures on ASIC and Arria 10 FPGA (ME) platforms. It shows that the \ac{rm} architecture consumes 63\% less gates than the \ac{mcm} one. 

The result given in Table~\ref{tab:synASIC} for FPGA platforms shows that the \ac{rm} architecture consumes 80\% less ALMs, 15\% less registers compared to the \ac{mcm}-based one. On the other hand, the total number of used DSP blocks is 32 which is 2\% of the FPGA total DSP blocks. These implementations could be further optimized for FPGAs to enhance the system frequency. However, it is estimated that the \ac{rm} architecture will always give better performance.

A fair comparison with other studies in the literature is quite difficult. Most studies focus on earlier version of VVC, and there are few ASIC-based designs for the VVC MTS. For comparison, Table \ref{comparison2D} lists the key performance of state-of-the-art ASIC and FPGA-based works, including the VVC-MTS related works~\cite{ledernier, 8698857, Can-mert, Kammoun}. Gate count is the logical calculation part and it can be seen from Table~\ref{comparison2D} that compared with implementations of Fan {\it et al.} \cite{ledernier} and Mert {\it et al.} \cite{Can-mert}, our solution has obvious advantages. We present a unified transform architecture that can realize IDCT-II/IDST-VII/IDCT-VIII for transform unit of order 4,8,16,32 and 64 with a fixed throughput of 2 pixels per cycle. Practically, up to 80.5\% of area can be reduced compared to \cite{ledernier} and up to 76.7\% compared to \cite{Can-mert}. In term of ALMs, we provide 92.7\% reduction compared to implementation proposed by Kammoun {\it et al.} in \cite{Kammoun}. However, currently, no implementation have been found for the IDCT-II of order 64. Although, \cite{ledernier} and \cite{Kammoun} supports 2D for all transform types, the transform could be achieved only up to the 32$\times$32 block size.

\section{Conclusion}
\label{sec:conclusion}

In this paper, two hardware implementations of 1D VVC-MTS module of are proposed. The two architectures are able to implement the VVC inverse transforms including block sizes from 4$\times$4 to 64$\times$64. Up the best of our knowledge, this is the first implementation that supports all VVC-MTS sizes. For our VVC/HEVC/AVC ASIC decoder, we adopted the \ac{rm} architecture since it enables a significant area saving. 

In the future, we aim to extend this architecture to support two dimension transform and include the Low-Frequency Non Separable Transform LFNST and quantization block within a unified module.

\newpage 
\bibliographystyle{IEEEbib}
\bibliography{strings}

\end{document}